\documentclass[prl, aps, twocolumn, groupedaddress,showpacs]{revtex4}
\usepackage[english]{babel} 
\usepackage[english]{layout}
\usepackage{amsmath,amsfonts,amssymb}
\usepackage{graphicx}
\usepackage{float}
\usepackage{color}
\usepackage{braket}
\usepackage{version}
\usepackage{array,multirow,makecell}

\begin{document}

\title{Effective Theory of Non-Adiabatic Quantum Evolution Based on the Quantum Geometric Tensor}
\author{O. Bleu, G. Malpuech, D. D. Solnyshkov}
\affiliation{Institut Pascal, PHOTON-N2, University Clermont Auvergne, CNRS, 4 avenue Blaise Pascal, 63178 Aubi\`{e}re Cedex, France.} 

\begin{abstract}
We study the role of the quantum geometric tensor (QGT) in the evolution of two-band quantum systems. We show that all its components play an important role on the extra phase acquired by a spinor and on the trajectory of an accelerated wavepacket in any realistic finite-duration experiment. While the adiabatic phase is determined by the Berry curvature (the imaginary part of the tensor), the non-adiabaticity is determined by the quantum metric (the  real part of the tensor).  We derive, for geodesic trajectories, (corresponding to acceleration from zero initial velocity), the semiclassical equations of motion with non-adiabatic corrections. The particular case of a planar microcavity in the strong coupling regime allows to extract the QGT components by direct light polarization measurements and to check their effects on the quantum evolution.

\end{abstract}

\maketitle

In 1984, Berry \cite{berry1984quantal} has shown that the quantum evolution in a parameter space leads to the accumulation of an extra phase in the wave function, the famous Berry phase (already known in optics as the Pancharatnam phase \cite{Pan1956}). Over the last decades, this concept and its generalization -- Berry curvature -- were understood to be among the most general in physics. For instance, the topological insulators \cite{Hasan2010} are classified by the Chern number\cite{Chern1946} -- an integer topological invariant obtained by integrating the Berry curvature over a complete energy band. Berry curvature  also  strongly affects the trajectory of an accelerated wave packet (WP), creating a lateral drift: an anomalous velocity, transverse to the acceleration. This anomalous velocity is at the origin of many crucial phenomena in physics such as the anomalous Hall effect (AHE) \cite{Luttinger1954,Sundaram1999,MacDonald2010}, the intrinsic Spin Hall effect for electrons\cite{Sinova2004} and light \cite{Bliokh2008,Ling2017}, or the Valley Hall effect \cite{Niu2007,mak2014valley} in Transition Metal Dichalcogenides (TMDs), the latter being a pillar of the emerging field called "valleytronics" \cite{schaibley2016valleytronics}.

However, Berry curvature is actually a part of a more general object: the quantum geometric tensor (QGT).
The gauge invariant QGT was introduced in 1980 by Provost and Vallee \cite{provost1980riemannian} as a part of a geometric approach to quantum mechanics \cite{Anandan1990}. The imaginary part of this Hermitian tensor corresponds to the Berry curvature  \cite{berry1984quantal,berry1989quantum}, whereas its real part defines a Riemannian metric, which allows to measure the distance between quantum states (Fubini-Study metric). Several recent theoretical works discuss the properties of this quantum metric for adiabatic quantum computation or to the study of phase transitions \cite{PhysRevLett.99.100603,PhysRevB.88.064304,PhysRevB.87.245103}.
In condensed matter, the real part of the QGT has been shown to be linked to the superfluid fraction of flat bands \cite{peotta2015superfluidity,PhysRevLett.117.045303}, to current noise in an insulator \cite{PhysRevB.87.245103}, to Lamb shift analog for exciton states in TMDs \cite{PhysRevLett.115.166802}, and to orbital magnetic susceptibility in Bloch bands \cite{Gao2015,Piechon2016}. 

The two parts of the QGT play complementary roles when the Hamiltonian of the system changes over time. The imaginary part (Berry curvature) defines the additional Berry phase in the adiabatic limit, while the real part (quantum distance) determines the non-adiabaticity (NA), which, in turn, brings a  correction to the Berry's formalism. 
NA in quantum systems has been studied extensively since the pioneering works of Landau \cite{Landau1932,Landau1932b}, Zener \cite{Zener1932}, Dykhne \cite{Dykhne1960}, and many others \cite{Landau1936,Stuckelberg1932,Davis1976}, concerning the regime where the NA is exponentially small, whereas configurations with power-law NA were generally considered as somewhat less interesting \cite{Dykhne1960,Landau3}. 
 The Landau quasi-classical formalism allows to calculate the final non-adiabatic fraction (transition probability) when the perturbation  smoothly vanishes at infinities. However, this approach cannot be applied to a simple yet important situation of a spin following a magnetic field rotating with a constant angular velocity, because the perturbation does not vanish.
Berry himself trusted that the NA is exponentially small \cite{berry1984quantal}, but that is not the case in the configuration he considered \cite{bliokh2008spin,oh2016singularity}, as we shall see below. Moreover, the NA changes during the evolution, and its final value (residual NA) is different from the maximal one. The Landau-Zener formalism allows to find only the former, while the latter is not exponentially small even if the evolution is perfectly smooth. In all these cases, the real part of the QGT allows to quantify the NA  and brings an important correction to the Berry phase.

In this work, we calculate the non-adiabatic corrections (NAC) for the phases and trajectories of  WPs for a finite-time quantum evolution beyond the Landau-Zener approximation, considering the important family of geodesic trajectories, corresponding to acceleration from zero initial velocity. We show that these NACs are quantitatively described by the real part of the QGT, whereas the adiabatic limit is described by the imaginary part (Berry phase). We propose an example of application of this formalism in a specific system: a planar microcavity \cite{Microcavities} in the strong coupling regime. We show that the use of such photonic structures allows, through simple light polarization measurements, a direct access to the components of the QGT in reciprocal space, providing an answer to an important problem of the recent years -- the direct measurement of Berry curvature and geometric quantities \cite{Onoda2004,Jotzu2014,Ozawa2014,Hafezi2014,Price2016,Flaschner2016,wimmer2017experimental,Montambaux2015,Montambaux2015b}. We consider a practical experimental situation showing how the real and imaginary part of the QGT control the AHE.

\emph{Rotation of a spin.}
The Bloch sphere represents the simplest 2-level system with Berry curvature: a spin, interacting with an applied magnetic field. Any 2-level Hamiltonian can be written as a superposition of Pauli matrices, and thus considered as an effective field acting on a pseudospin.

\begin{figure}[tbp]
 \begin{center}
 \includegraphics[scale=0.42]{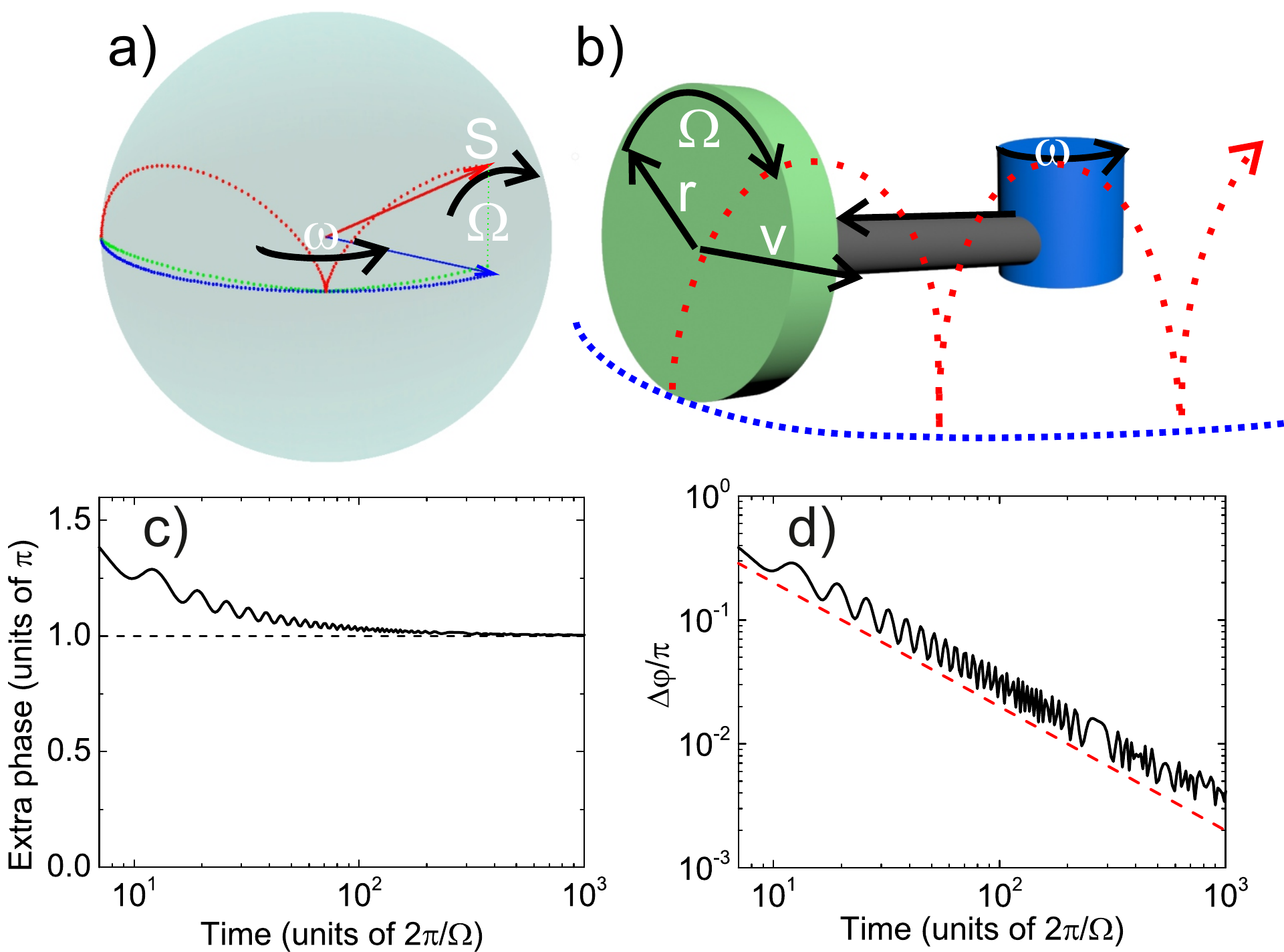}
 \caption{\label{figSphere}(Color online)(a) Bloch sphere with the spin (red arrow) and the magnetic field $\Omega$ (blue arrow), adiabatic trajectory (blue) and real trajectory (red dashed line); (b) Mechanical analog: "adiabatic" trajectory of an infinitely small wheel (blue), cycloid trajectory of a point on a wheel (red). $\Omega$ - wheel rotation frequency, $v$ - wheel velocity, $\omega$ - shaft center rotation; (c) Total extra phase for one full spinor rotation as a function of the rotation time; (d)  Deviation from the adiabatic Berry phase: numerical calculation (black) and analytical correction exhibiting $1/T$ decay (red dashed).}
  \end{center}
 \end{figure}

A spin, which follows a slowly rotating magnetic field, is never perfectly aligned with it, and thus it exhibits fast precession (with frequency $\Omega$) about the magnetic field together with the slow rotation (freq. $\omega$) of both of them in the azimuthal plane (see Fig. \ref{figSphere}(a)). This behavior is similar to the rotation of a small wheel attached to a long shaft (Fig. \ref{figSphere}(b)): the wheel, rotating around its axis with the angular frequency $\Omega$, at the same time rotates with the frequency $\omega$ around the shaft fixation point. For both the spin and the wheel, there is an important rotational energy associated with the large frequency $\Omega$, but another part of the energy is associated with the circular motion $\omega$. Nobody could think of neglecting the kinetic energy of the wheel's motion $mv^2/2$. However, the energy of the spin's slow rotation encoded in the Berry phase has been less evident to see. It can be obtained by applying the energy operator $\hat{E}=i\hbar\partial/\partial t$ to the rotating spinor $\psi(t)=1/\sqrt{2}(e^{-i\omega t},1)^T e^{i\Omega t/2}$ (valid in the limit $\omega\to 0$), which gives $\langle \hat{E}\rangle =-\hbar\Omega/2+\hbar\omega/2$. The first term in this expression is the usual energy of the spin in the magnetic field ("dynamical phase"), and the second is the energy associated with the Berry phase which appears because of the time dependence of the spinor. For the time $T=2\pi/\omega$ of one full rotation of the field it gives $\gamma_B=\hbar\omega T/\hbar=\pi$. Taking into account only the interaction of the spin with the magnetic field $-\hbar\mathbf{\Omega}\mathbf{S}/2$ is like going into the reference frame of the disk: we should not forget that this frame is moving.
One can then take a derivative over the parameters of the wavefunction (WF), to get rid of the explicit time dependence $i\left\langle {\psi }
 \mathrel{\left | {\vphantom {\psi  {\partial \psi /\partial t}}}
 \right. \kern-\nulldelimiterspace}
 {{\partial \psi /\partial t}} \right\rangle  = i\left\langle {\psi }
 \mathrel{\left | {\vphantom {\psi  {\partial \psi /\partial q}}}
 \right. \kern-\nulldelimiterspace}
 {{\partial \psi /\partial \varphi}} \right\rangle \partial \varphi/dt$.

Because of the finite experiment duration, the spin does not perfectly follow the field and gets out of the azimuthal plane, tracing a cycloidal trajectory. The corresponding WF can be written as
\begin{equation}
\psi \left( t \right) = \left( {\begin{array}{*{20}{c}}
{\cos \frac{{\theta \left( t \right)}}{2}{e^{ - i\omega t}}}\\
{\sin \frac{{\theta \left( t \right)}}{2}}
\end{array}} \right){e^{i\frac{{\Omega \cos \xi \left( t \right)}}{2}t}}
\label{eqspinor}
\end{equation}
where $\theta$ is the polar angle  and $\xi$ is the angle between the field and the spin. Averaging this expression over precession time allows obtaining the  correction to the energy. The average value of $\theta$ for the cycloidal trajectory of Fig. \ref{figSphere}(a) corresponds to the equilibrium polar angle $\theta=\pi/2-\omega/\Omega$, at which the spin can follow the field without precession \cite{suppl,Halperin2007,oh2016singularity}, giving \cite{suppl} $E=-\hbar\Omega/2+\hbar\omega/2(1+2\omega/\Omega)$, and a final extra phase $\gamma=\pi(1+2\omega/\Omega)$ or  $\Delta\gamma/\gamma_B=4\pi/(\Omega T)$ for one full rotation time $T$.

The total extra phase after one full rotation of the magnetic field from the numerical solution of the Schr\"odinger equation is plotted in Fig. \ref{figSphere}(c) as a function of the rotation duration $T$ measured in units of precession periods $2\pi/\Omega$ (equivalent to the frequency ratio $\Omega/\omega$). Larger $T$ means slower rotation and the adiabatic limit corresponds to $T\to\infty$ or $\omega/\Omega\to 0$. We see that the extra phase indeed converges to the value $\pi$, but the correction is not negligible: $\Delta\gamma/\gamma_B>30\%$ for $\omega>\Omega/10$. The difference between the exact extra phase and the adiabatical value of $\pi$ is shown in a Log-Log plot on Fig. \ref{figSphere}(d), again as a function of $T$ (black curve).  We see that instead of being exponentially small, this correction decreases only as $1/T$. The analytical NAC, $\Delta\gamma/\gamma_B=4\pi/(\Omega T)$ (red curve), fits the exact result very well \cite{suppl}.

\emph{Quantum Geometric Tensor.} The QGT allows to generalize the above development to an arbitrary parameter space and to unite both contributions to the extra phase acquired by the WF in a single mathematical entity. In general, a metric tensor $g_{ij}$ determines how the distance $ds$ between two infinitesimally separated points depends on the difference of their coordinates $\lambda_i$:
\begin{equation}
ds^2=g_{ij}d\lambda_id\lambda_j
\end{equation}
In the space of quantum-mechanical eigenstates, the distance is measured by the Fubini-Study metric, determined by the WF overlap $ds^2=1-|\braket{\psi(\mathbf{\lambda})|\psi(\mathbf{\lambda}+\delta\mathbf{\lambda})}|^2$. Minimal distance $ds=0$ corresponds to a maximal overlap of $1$, while maximal distance $ds=1$ corresponds to orthogonal states. At each point of the Hilbert space, the metric is thus determined by the WF $\psi(\mathbf{\lambda})$, and the corresponding metric tensor is defined as the real part of \cite{provost1980riemannian}:
\begin{equation}
{T_{ij}} = \left\langle {{\frac{\partial }{{\partial {\lambda_i}}}\psi }}
 \mathrel{\left | {\vphantom {{\frac{\partial }{{\partial {\lambda_i}}}\psi } {\frac{\partial }{{\partial {\lambda_j}}}\psi}}}
 \right. \kern-\nulldelimiterspace}
 {{\frac{\partial }{{\partial {\lambda_j}}}\psi}} \right\rangle  - \left\langle {{\frac{\partial }{{\partial {\lambda_i}}}\psi}}
 \mathrel{\left | {\vphantom {{\frac{\partial }{{\partial {\lambda_i}}}\psi} \psi}}
 \right. \kern-\nulldelimiterspace}
 {\psi} \right\rangle \left\langle {\psi}
 \mathrel{\left | {\vphantom {\psi  {\frac{\partial }{{\partial {\lambda_j}}}\psi}}}
 \right. \kern-\nulldelimiterspace}
 {{\frac{\partial }{{\partial {\lambda_j}}}\psi}} \right\rangle 
\end{equation}
where $\psi$ is the WF, $\lambda_i$ and $\lambda_j$ are the coordinates in the parameter space. 
Later, it was understood that the imaginary part of this tensor is the Berry curvature \cite{berry1989quantum}:
\begin{equation}
|\mathbf{B}|=2\Im[T_{ij}]=\Im[\nabla_\mathbf{\lambda}\times\braket{\psi(\mathbf{\lambda})|\nabla_\mathbf{\lambda}\psi(\mathbf{\lambda})}]
\end{equation}
which determines the Berry phase for a closed path in the parameter space.

Both components of the QGT contribute to the phase of the WF in any finite-duration experiment: Berry curvature determines the adiabatic value for the phase, while the quantum metric allows to determine a correction due to the NA. The average NA fraction (fraction of the excited state in the WF) for a spin on the Bloch sphere can be found as $f_{NA,eq}=\omega^2/4\Omega^2$ \cite{suppl}, which is generalized by using $
\omega \left( \lambda \right)=2ds/dt=2\sqrt {{g_{\lambda\lambda}}\left( \lambda \right)} d\lambda/dt$
and $\Omega=\Omega(\lambda)$:
\begin{equation}
{f_{NA,eq}}\left( \lambda \right) = \frac{{{g_{\lambda\lambda}}}}{{{\Omega^2}}}\left(\frac{d\lambda}{dt}\right)^2
\end{equation}
It determines the corrections to the Berry phase and also to the AHE trajectory for finite-duration experiments in a general parameter space.

\emph{QGT and WP trajectory.} 
Berry curvature has been shown to affect the trajectory of accelerated WPs, creating an anomalous velocity contribution in the AHE \cite{Luttinger1954,Sundaram1999}. The semiclassical equations of motion for the center of mass of a quantum WP in presence of Berry curvature can be derived using the Lagrangian formalism \cite{Chang1995,Sundaram1999,Culcer2005,RevModPhys.82.1959,Chang2008}):
\begin{equation}
\label{sem}
\hbar\frac{\partial\textbf{k}}{\partial t}=\textbf{F}, \quad
\hbar \frac{\partial\textbf{r}}{\partial t}= \frac{\partial \epsilon}{\partial \textbf{k}}  - \hbar \frac{\partial\textbf{k}}{\partial t} \times \textbf{B} 
\end{equation}
where $\epsilon$ is the energy dispersion, $\textbf{B}(k)$ is the Berry curvature and $\mathbf{F}$ is an external conservative force, accelerating the wave-packet. For charged particles, $\mathbf{F}$ is an electric force. Magnetic forces, known to affect the magnetic susceptibility \cite{Ceresoli2006,Gao2015,Piechon2016}, are not the subject of the present work. Different types of corrections to these equations have been considered in the past \cite{Mohrbach2010,Gao2014,Gao2016}.
Non-adiabatic corrections account for the fact that the WF is a superposition of two eigenstates $\psi=f_0\psi_0+f_1\psi_1$ (where $|f_1|^2=f_{NA}$ found above). Their respective energies contribute both to the first term: $\tilde \epsilon\left( k \right) = {|f_0|^2}{\epsilon_0}\left( k \right) + {|f_1|^2}{\epsilon_1}\left( k \right)$, ultimately providing a second-order correction to the group velocity. Other NA corrections concern the second term, and, in a general case, the first-order corrections found in \cite{Gao2014} should dominate. 

Along geodesic lines, which is actually the most important case, corresponding to acceleration from $v=0$ under a constant force $\mathbf{F}$ (the configuration of Hall effect), all first-order and second-order corrections cancel, except one. This single correction  appears because the metric along the true trajectory of $\psi$ is not the same as the one along equator of the Bloch sphere (followed by the eigenstates $\psi_0$ and $\psi_1$). Indeed:
\begin{equation}
\left\langle {\psi }
 \mathrel{\left | {\vphantom {\psi  {\frac{d}{{dt}}\psi }}}
 \right. \kern-\nulldelimiterspace}
 {{\frac{d}{{dt}}\psi }} \right\rangle  = \left\langle {\psi }
 \mathrel{\left | {\vphantom {\psi  {\frac{d}{{ds}}\psi }}}
 \right. \kern-\nulldelimiterspace}
 {{\frac{d}{{ds_s}}\psi }} \right\rangle \frac{{ds_s}}{{dt}} = \left\langle {\psi }
 \mathrel{\left | {\vphantom {\psi  {\frac{d}{{d\varphi }}\psi }}}
 \right. \kern-\nulldelimiterspace}
 {{\frac{d}{{d\varphi }}\psi }} \right\rangle \frac{{d\varphi }}{{ds_s}}\frac{{ds_s}}{{dt}}
\end{equation}
  where $d\varphi/ds_s=1/\sqrt{g_{\varphi\varphi}}=1/r\sin\theta$.
 Now, we can write $\psi$ and the Berry connection on the basis of the eigenstates (which are on the equator, where $ds_s=d\varphi$):
 \begin{equation}
\frac{1}{{\sin \theta }}\left\langle {{{f_i}{\psi _i}}}
 \mathrel{\left | {\vphantom {{{f_i}{\psi _i}} {\frac{d}{{d\varphi }}{f_i}{\psi _i}}}}
 \right. \kern-\nulldelimiterspace}
 {{\frac{d}{{d\varphi }}{f_i}{\psi _i}}} \right\rangle \frac{{ds_s}}{{dt}} = \frac{1}{{\sin \theta }}\left\langle {{{f_i}{\psi _i}}}
 \mathrel{\left | {\vphantom {{{f_i}{\psi _i}} {\frac{d}{{dt}}{f_i}{\psi _i}}}}
 \right. \kern-\nulldelimiterspace}
 {{\frac{d}{{dt}}{f_i}{\psi _i}}} \right\rangle  
 \end{equation}
The Berry connection above involves intra- and inter-band terms \cite{Gao2014}. For the Berry curvature appearing in the AHE, the intraband terms add up to $1$, while the inter-band terms cancel out, giving simply $B=B_0/\sin\theta$, where $B_0$ is the Berry curvature of the instantaneous eigenstate $\psi_0$. The Lagrangian formalism provides \cite{suppl} the corrected equation for the trajectory: 
\begin{equation}
\hbar \frac{\partial\textbf{r}}{\partial t}= \frac{\partial \tilde \epsilon}{\partial \textbf{k}}  - \hbar \frac{\partial\textbf{k}}{\partial t} \times 2\Im\left[\textbf{T}_{k\phi}\right]\left(1+2\frac{T_{kk}}{\Omega^2}\left(\frac{\partial\textbf{k}}{\partial t}\right)^2\right)
\label{semiclassical}
\end{equation}

This equation is the main result of our manuscript. It shows that the anomalous velocity is a sum of the adiabatic value (as in Eq. \eqref{sem}) and a NAC (the second term in the parenthesis). We stress that this equation is only valid when the field follows a geodesic trajectory in the parameter space, as in the AHE.
 In such case, while the renormalized energy $\tilde \epsilon$ brings second-order corrections to the acceleration in the direction of the force, the anomalous velocity only includes the correction from the variation of the metric due to the NA, because the other first and second-order corrections to this term cancel out. 
%In such case, all first-order corrections and all additional second-order corrections cancel out, and the anomalous velocity only includes the correction from the variation of the metric due to the NA. However, the renormalized energy $\tilde \epsilon$ brings second-order corrections to the acceleration in the direction of the force.

\begin{figure}[tbp]
 \begin{center}
 \includegraphics[scale=0.35]{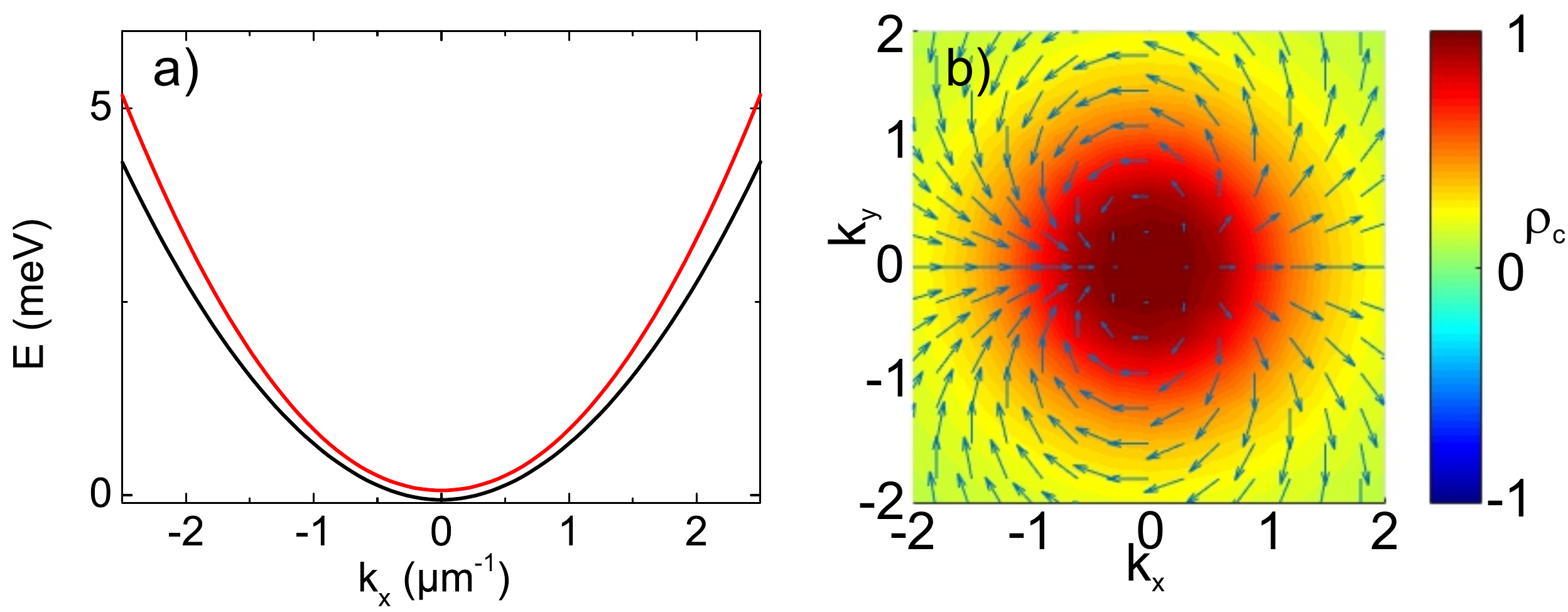}
 \caption{(a) LPB split by Zeeman field and TE-TM SOC; (b) Pseudospin texture of the lower eigenstate: in-plane pseudospin projection (arrows) and $S_Z$ (color).}
  \end{center}
 \end{figure}

\emph{QGT in a planar cavity.} 
Exploring the whole Bloch sphere requires all 3 components of the effective magnetic field. If one deals with light, it means controlling the splittings between linear and circular polarizations. This is why we have chosen a model system consisting of a microcavity in the strong coupling regime \cite{Microcavities}, where the polariton modes appear from exciton and photon resonances. The photonic fraction provides a $\beta k^2$ in-plane spin-orbit coupling (SOC) due to the TE-TM splitting \cite{Panzarini99,Kavokin2005}, while the exciton mode provides the Zeeman splitting $\Delta$ \cite{Tignon1995,Deveaud2015} (under applied magnetic field or thanks to spin-anisotropic interactions \cite{Vladimirova2010}). A remarkable feature of this scheme is that the pseudo-spin can be easily measured via the polarization of light \cite{suppl}.

We begin with the parabolic spinor Hamiltonian of the lower polariton branch (LPB) of a planar cavity:
\begin{eqnarray}
H_0=\begin{pmatrix}
\frac{\hbar^2k^2}{2m^*}+\Delta&\beta k^2 e^{2i\phi} \\
\beta k^2 e^{-2i\phi}& \frac{\hbar^2k^2}{2m^*}-\Delta
\end{pmatrix} 
\end{eqnarray}
The eigenvalues of this Hamiltonian, shown in Fig. 2(a), are:
\begin{equation}
\epsilon_{\pm}(k)=\frac{\hbar^2k^2}{2m^*}\pm\sqrt{(\Delta^2+\beta^2k^4)}
\label{energies}
\end{equation}
We have used $\Delta=60$ $\mu$eV and $\beta=0.14$ meV/$\mu$m$^{-2}$. While the system clearly shows no gap because of the positive mass of both branches, and therefore is not an example of topological insulator, it nevertheless exhibits a non-zero Berry curvature, reflected by the pseudospin texture (Fig. 2(b)). This dipolar pseudospin texture is also typical for bilayer graphene around K points \cite{macdonald2012pseudospin}, where quadratic degeneracies can be opened by a bias voltage \cite{McCann2006,GeimCastro2007}. 
We compute analytically the QGT for the lower eigenstate in polar coordinates ($k$,$\phi$):
\begin{eqnarray} \label{QGT}
g_{kk}=\frac{\Delta^2 k^2\beta^2}{(\Delta^2+\beta^2k^4)^2}, \quad
g_{\phi\phi}=\frac{k^2\beta^2}{\Delta^2+\beta^2k^4} \\
g_{k\phi}=g_{\phi k}=0, \quad
\mathbf{B}=\frac{2\Delta k^2\beta^2}{(\Delta^2+k^4\beta^2)^{3/2}}\mathbf{e_Z} \nonumber
\end{eqnarray}
These are plotted as solid curves in Fig. 3. Because of the $k^2$ dependence of the TE-TM SOC, the form of the Berry curvature is different from the one of Rashba SOC \cite{RevModPhys.82.1959, MacDonald2010} (with maximum at $k=0$) and similar to the one of bilayer graphene \cite{Macdonald2011}.

A very interesting opportunity to measure these QGT components directly is offered by the radiative states of photonic systems which allow to access all pseudospin components $\mathbf{S}$ via polarization:
\begin{eqnarray}
g_{kk}&=& \frac{1}{4}\frac{{{{\left( {\frac{\partial }{{\partial k}}{S_z}(k)} \right)}^2}}}{{1 - {S_z}{{(k)}^2}}}\\
{g_{\phi \phi }} &=& \frac{1}{{4{k^2}}}{\left( {\frac{{\frac{\partial }{{\partial \phi }}\left( {\frac{{{S_y}}}{{{S_x}}}} \right)}}{{1 + {{\left( {\frac{{{S_y}}}{{{S_x}}}} \right)}^2}}}} \right)^2}\left( {1 - S_z^2} \right)\\
|\textbf{B}|& =& \frac{1}{{2k}}{\left( {\frac{{\frac{\partial }{{\partial \phi }}\left( {\frac{{{S_y}}}{{{S_x}}}} \right)}}{{1 + {{\left( {\frac{{{S_y}}}{{{S_x}}}} \right)}^2}}}} \right)^2}\frac{{\partial {S_z}}}{{\partial k}}
\end{eqnarray}
To demonstrate that the QGT components including the Berry curvature can indeed be extracted from a realistic experiment, we perform a numerical simulation using a 2D spinor Schr\"odinger equation written for LPB in the parabolic approximation:
\begin{eqnarray}\label{schro}
& i\hbar \frac{{\partial \psi _ \pm  }}
{{\partial t}}  =  - \frac{{\hbar ^2 }}
{{2m}}\Delta \psi _ \pm  - \frac{{i\hbar }}
{{2\tau }}\psi _ \pm  +\Delta\psi_\pm\\
& + \beta {\left( {\frac{\partial }{{\partial x}} \mp i\frac{\partial }{{\partial y}}} \right)^2}{\psi _ \mp } + \hat{P} \nonumber 
\end{eqnarray}
where ${\psi_+(\mathbf{r},t), \psi_-(\mathbf{r},t)}$ are the two circular components, $m=5\times10^{-5}m_{el}$ is the polariton mass, $\tau=30$ ps the lifetime, $\hat{P}$ is the pump operator which in this case represents uncorrelated noise describing the spontaneous scattering under non-resonant pumping of the exciton reservoir. The results of the extraction are presented in Fig. 3 as dashed curves, whose excellent agreement with the solid lines obtained from Eq.~\eqref{QGT} confirms the validity of this method.

 \begin{figure}[tbp]
 \begin{center}
 \includegraphics[scale=0.29]{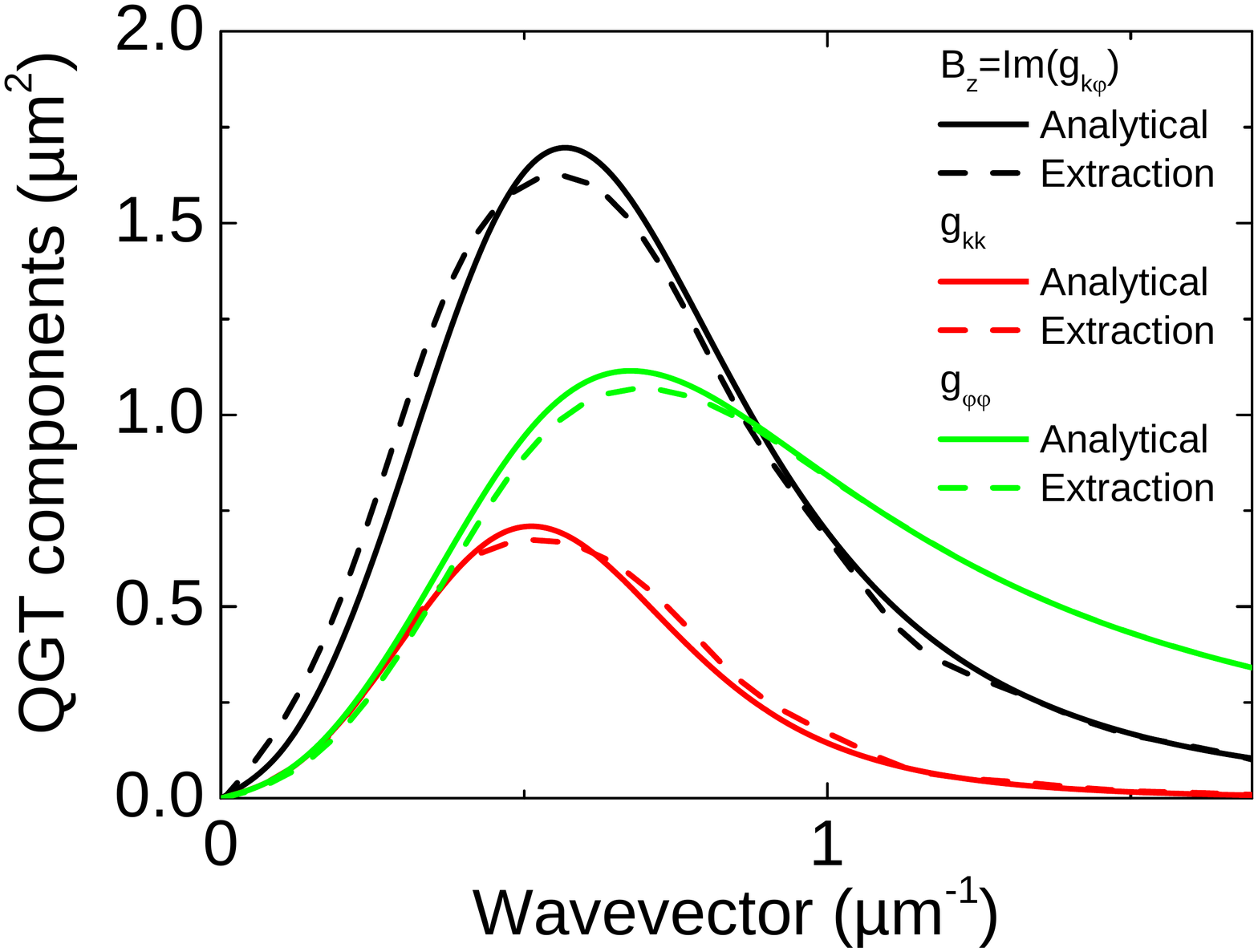}
 \caption{(Color online) QGT components: $B_Z$ (black); $g_{kk}$ (red) and $g_{\phi\phi}$ (green) calculated analytically (solid lines)  and extracted from numerical experiment (dashed lines).}
  \end{center}
 \end{figure}

 \begin{figure}[tbp]
 \begin{center}
 \includegraphics[scale=0.33]{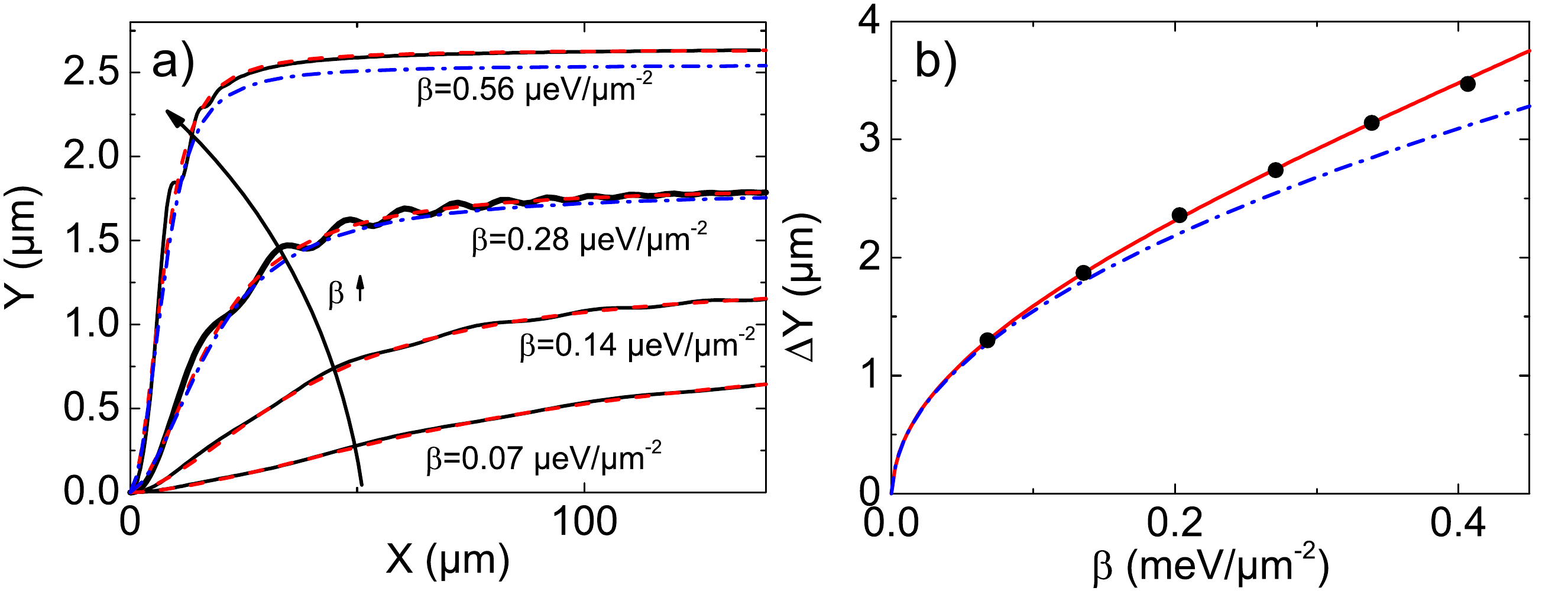}
 \caption{(Color online) (a) WP trajectories in real space: numerical (black) and analytical (red dashed, uncorrected - blue dash-dotted) for 3 values of TE-TM SOC $\beta$ ($\Delta=0.06$ meV); (b) Final lateral shift as a function of $\beta$: adiabatic (blue dash-dotted line), corrected (red solid line), and numerical (black dots). Here, $\Delta=0.03$ meV. }
  \end{center}
 \end{figure}

Figure 4(a) shows  the trajectories of polariton WP accelerated in a microcavity by a realistic wedge $U(x)=-Fx$, where $F=1~$meV/$128~\mu$m for 3 different values of $\beta$. The red-dashed curves are analytically calculated using the equation \eqref{semiclassical} and are in excellent agreement with direct numerical solution of the Schr\"odinger equation \eqref{schro} (black curves).  The NA fraction can be extracted experimentally by doing polarization measurements: $f_{NA}=S_Y^2$ (see \cite{suppl} for details).  
The blue dotted curve shows the trajectory without the correction. The difference becomes more important for higher gradients. Fig. 4(b) shows the final lateral shift $\Delta Y$ as a function of $\beta$: adiabatic ($\Delta Y=\sqrt{\beta}\Gamma^2(3/4)/\sqrt{\Delta\pi}$ - blue dotted \cite{suppl}) and corrected (red) curves, as well as results of simulations (black dots). Numerical results are much better fitted by the theory including the NAC.

To conclude, we derive a new correction to the semiclassical equations of motion of an accelerated WP on geodesic trajectories in two-band systems appearing in any realistic finite-duration experiment. While the adiabatic limit is determined by the Berry curvature, the NAC is determined by the quantum metric.  The particular case of a planar microcavity in strong coupling regime allows to extract the QGT components by direct measurements and to check their effects on the quantum evolution.

\begin{acknowledgments}
We thank F. Piechon and J.N. Fuchs for useful discussions. We acknowledge the support of the project "Quantum Fluids of Light"  (ANR-16-CE30-0021), of the ANR Labex Ganex (ANR-11-LABX-0014), and of the ANR Labex IMobS3 (ANR-10-LABX-16-01).
\end{acknowledgments}

 \bibliography{biblio} 

\section{Supplemental material} 
 
In Section I, we derive the non-adiabatic fraction for a two-level system in a magnetic field rotating with a constant angular velocity. In Section II, we analyze the effects of acceleration. In Section III, we present the computation of the average correction to the energy due to the real cycloidal trajectory of the spin. In Section IV, we derive the semiclassical equations of motion for a wavepacket on a geodesic trajectory using the effective Lagrangian formalism. Section V shows that the previously found first-order corrections cancel on the geodesics. In Section VI, we provide details on the  non-adiabatic evolution in the particular case of polaritons. We describe the experimental measurement of the non-adiabaticity in such systems in Section VII. In Section VIII, we present the analytical expressions for the polariton anomalous Hall drift obtained using the corrected semiclassical equations presented in the main text. Finally, a movie (available as a separate file) of a numerical simulation of the polariton anomalous Hall effect is discussed in Section IX.

\subsection{Equations for spin dynamics in a rotating field}
\begin{figure}[tbp]
\includegraphics[scale=0.5]{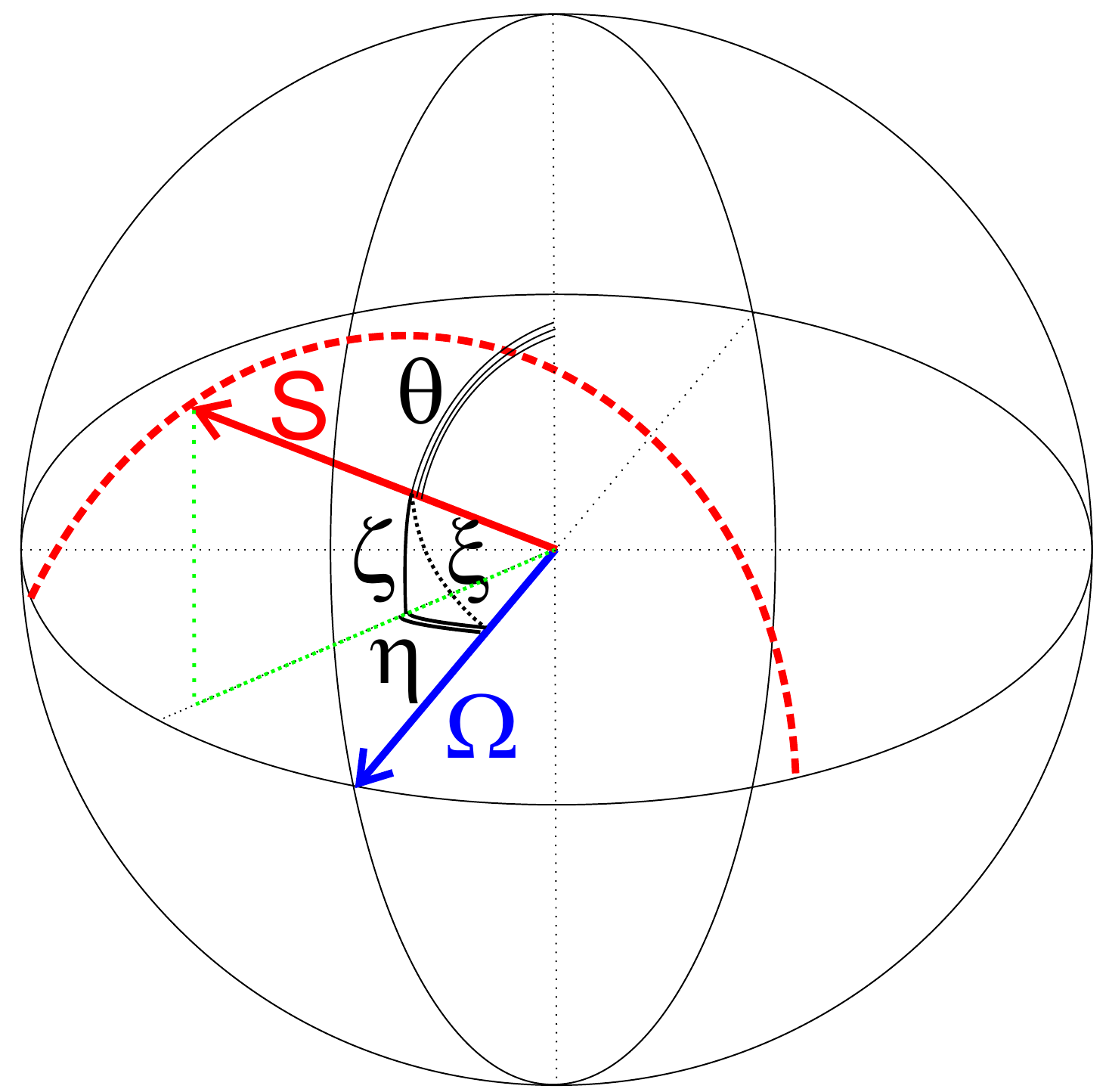}
\caption{\label{figscheme} (Color online) The spin vector $\mathbf{S}$ (red), its projection on the equatorial plane (green), and the magnetic field $\mathbf{\Omega}$ (blue) on the Bloch sphere, with the angles used in the supplemental material and in the main text: $\theta$ (polar angle of $\mathbf{S}$, $\zeta$ (latitude of $\mathbf{S}$), $\eta$ (difference between the longitudes of $\mathbf{S}$ and $\mathbf{\Omega}$), and $\xi$ (angle between $\mathbf{S}$ and $\mathbf{\Omega}$). Red dashed line shows the cycloidal spin trajectory. }
\end{figure}

We begin with the precession equation for the spin dynamics, which can be obtained from the spinor Schrodinger equation. This equation reads:

\begin{equation}
\frac{{d{\bf{S}}}}{{dt}} = {\bf{\Omega }} \times {\bf{S}}
\label{precess}
\end{equation}
Here, $\mathbf{\Omega}(t)$ is the magnetic field vector, which can vary both in direction and magnitude in the general case, and $\mathbf{S}$ is the spin vector. To be specific, we consider the rotation of the magnetic field in the equatorial plane (a geodesic trajectory on a sphere). The scheme of the Bloch sphere with the spin and the magnetic field and their relative angles is shown in Fig. S\ref{figscheme}.

Since we are going to study small deviations from the adiabatic regime, the motion of the spin vector is limited to small angles close to the equator of the sphere. We can rewrite the equations using two new variables: the angle between the spin and the equatorial plane $\zeta$ (latitude), and the azimuthal angle between the spin projection on this plane and the magnetic field $\eta$. These equations read:

\begin{equation}
\begin{array}{l}
\frac{{d\zeta }}{{dt}} =  - \Omega \eta \\
\frac{{d\eta }}{{dt}} = \Omega \left( {\zeta  - \frac{\omega }{\Omega }} \right)
\end{array}
\label{dyneq}
\end{equation}
Here, $\Omega$ is the magnitude of the magnetic field, while $\omega$ is its angular velocity: $\omega=d\phi/dt$ ($\phi$ is the azimuthal angle of the magnetic field).
We see that these equations are similar to the Hamilton's equations for a harmonic oscillator in rescaled coordinates ($\eta$ plays the role of coordinate $q$ and $\zeta$ -- the role of momentum $p$), but with an extra term $d\eta/dt=-\omega$, corresponding to a constant velocity of the harmonic oscillator, meaning that we consider it in a moving reference frame, or that the oscillator itself is moving in our fixed reference frame. This falls very well within the analogy with a rotating wheel considered in the main text.

 First of all, we can find a stationary solution: 

\begin{equation}
\zeta_{eq}=\frac{\omega}{\Omega}
\end{equation}
It means that if the magnetic field rotates with a constant angular velocity $\omega$, the spin is able to follow it with the same angular velocity \emph{if} it is deviated from the equatorial plane towards the pole by $\omega/\Omega$. In this case, $\eta=0$. This solution is very important, because it allows us to estimate the non-adiabaticity in a general case considered below. The fraction of the excited state in the wavefunction is given by:

\begin{equation}
{f_{NA}} ={\left| {\left\langle {{\begin{array}{*{20}{c}}
{\cos \left( {\frac{1}{2}\left( {\frac{\pi }{2} - \theta } \right)} \right)}\\
{\sin \left( {\frac{1}{2}\left( {\frac{\pi }{2} - \theta } \right)} \right)}
\end{array}}}
 \mathrel{\left | {\vphantom {{\begin{array}{*{20}{c}}
{\cos \left( {\frac{1}{2}\left( {\frac{\pi }{2} - \theta } \right)} \right)}\\
{\sin \left( {\frac{1}{2}\left( {\frac{\pi }{2} - \theta } \right)} \right)}
\end{array}} {\begin{array}{*{20}{c}}
1/\sqrt{2}\\
{ - 1/\sqrt{2}}
\end{array}}}}
 \right. \kern-\nulldelimiterspace}
 {{\begin{array}{*{20}{c}}
1/\sqrt{2}\\
{ - 1/\sqrt{2}}
\end{array}}} \right\rangle } \right|^2} = \frac{{{\zeta ^2}}}{4}
\end{equation}
This fraction is complementary to fidelity $F=|\langle \psi|\psi_0\rangle|$, where $\psi_0$ is the ground state in which the system is expected to remain:
\begin{equation}
f_{NA}+F^2=1
\end{equation}

If at $t=0$ the spin is not in the position corresponding to the stationary solution (the equilibrium point $\zeta=\zeta_{eq}$, $\eta=0$), it will precess about the equilibrium position, exactly as the harmonic oscillator with some initial displacement oscillates around the equilibrium point. Since the system of equations is the same as for the harmonic oscillator, its solutions belong to the same family of curves in the phase space (represented here by $\theta$,$\eta$): for small angles, these are circular orbits around the equilibrium position, where the radius of the circle is determined by the energy.
For example, if at $t=0$ the spin is in equatorial plane, it will oscillate around the equilibrium position $\zeta_{eq}=\omega/\Omega$, that is, between $\zeta=0$ and $\zeta=2\zeta_{eq}$. Its average non-adiabaticity will be given by

\begin{equation}
f_{NA}=\frac{\omega^2}{4\Omega^2}
\end{equation}

\subsection{Angular acceleration and residual non-adiabaticity}

But what if the angular velocity of the rotation of the magnetic field is changing? Changing the ratio $\omega/\Omega$ corresponds to modification of the equilibrium position of the harmonic oscillator, which is possible under the action of a force. This is not a periodic driving force, contrary to the well-known driven damped oscillator problem, but rather a slowly-varying or stepwise-constant force. 

To make this force appear in the equations, let us introduce a new variable $\zeta'=\zeta-\omega/\Omega$. It allows to rewrite the system of equations as follows:

\begin{equation}
\begin{array}{l}
\frac{{d\zeta '}}{{dt}} =  - \Omega \eta  - \frac{d}{{dt}}\frac{\omega }{\Omega }\\
\frac{{d\eta }}{{dt}} = \Omega \zeta '
\end{array}
\end{equation}
We see that now the term $d(\omega/\Omega)/dt$ appears in the equation for the momentum derivative, and therefore it indeed plays the role of a force, which determines the equilibrium position. Of course, if a harmonic oscillator is moving (as a whole) and then suddenly stops, or is not moving and then suddenly set into motion, this sudden change excites it and increases the amplitude of its oscillations.

Since the system is not damped, these oscillations cannot decay, and if the angular velocity of the magnetic field is slowly reduced to 0, they will remain and provide a residual non-adiabaticity, which is not exponentially small, but is instead proportional to the derivative $d(\omega/\Omega)/dt$. The fact that if a $n$-th order derivative of the perturbation experiences a jump, the final transition probability (which is the residual non-adiabaticity in our terms) is not exponential, but proportional to the height of this jump power $n$, has been obtained in the 30ies within the approach of Landau \cite{Landau3} and is quite well-known, although considered as less interesting than the exponentially small transition probability for a smooth perturbation.

 In our case, the residual non-adiabaticity will not play an important role, because it corresponds to oscillations around the equilibrium position. The contribution of these oscillations averages out over time and does not contribute to the wavepacket trajectory (see further sections).
 
 On the other hand, the equilibrium position $\zeta_{eq}$ gives a non-vanishing contribution to the final phase and to the wavepacket trajectory via Berry curvature. It is therefore more important in our case to calculate the equilibrium non-adiabaticity as a function of time and to take into account its effects, than to study the residual non-adiabaticity (final transition probability), contrary to the cases considered in the previous works.

Finally, we note that the difference between the exact extra phase (black) and the analytically found solution (red) in Fig. 1d of the main text might be due to the fact that the length of the non-adiabatic cycloidal trajectory is different from the length of the trajectory of the magnetic field \cite{oh2016singularity}.

\subsection{Averaging for the spinor}
In the main text, we calculate the average correction to the energy due to the deviation of $\mathbf{S}$ from $\mathbf{\Omega}$. There are two contributions to the energy correction: one from the $\cos\theta/2$ term in the spinor (that is, from the non-zero latitude of the spin), which modifies directly the Berry phase, and one from the magnetic energy term $-\hbar\mathbf{\Omega}\mathbf{S}$, which can also be integrated into a correction for the Berry phase, as we shall see.

The energy operator applied on the spinor gives $\hbar\omega\cos^2\theta(t)/2$, which should be averaged over one precession period $2\pi/\Omega$. The time dependence of $\theta(t)$ is found from the equations for the spin dynamics: it is simply a harmonic oscillation.
\begin{equation}
\theta \left( t \right) = \frac{\pi }{2} - \frac{{1 - \cos \Omega t}}{2}\frac{\omega }{\Omega }
\end{equation}
The averaged contribution to the energy reads:
\begin{equation}
\frac{{\hbar \omega }}{{2\pi /\Omega }}\int\limits_0^{2\pi /\Omega } {{{\cos }^2}\left( {\frac{{\frac{\pi }{2} - \frac{{1 - \cos \Omega t}}{2}\frac{\omega }{\Omega }}}{2}} \right)} dt = \frac{{\hbar \omega }}{2}\left( {1 + \frac{\omega }{{\Omega }}} \right)
\end{equation}
This is the Berry phase plus a part of the non-adiabatic correction.

The magnetic energy also has to be averaged over time. The angle $\xi(t)$ between $\mathbf{S}$ and $\mathbf{\Omega}$ is given by 
\begin{equation}
\xi \left( t \right) = \frac{\omega }{\Omega }\sqrt {{{\left( {1 + \cos \Omega t} \right)}^2} + {{\sin }^2}\Omega t}
\end{equation}
The corresponding contribution to the energy reads
\begin{equation}
 - \frac{{\hbar \Omega }}{2}\cos \xi  \approx  - \frac{{\hbar \Omega }}{2}\left( {1 - \frac{{{\xi ^2}}}{2}} \right)
 \end{equation}
The averaging gives:
\begin{equation}
\begin{array}{l}
 - \frac{{\hbar \Omega }}{2}\left( {1 - \frac{1}{2}\frac{{{\omega ^2}}}{{{\Omega ^2}}}\frac{1}{{2\pi /\Omega }}\int\limits_0^{2\pi /\Omega } {\left( {{{\left( {1 + \cos \Omega t} \right)}^2} + {{\sin }^2}\Omega t} \right)dt} } \right) = \\
 - \frac{{\hbar \Omega }}{2} + \frac{{\hbar \omega }}{2}\frac{\omega }{\Omega }
\end{array}
\end{equation}

Finally, the energy can be written as:
\begin{equation}
\bar E =  - \frac{{\hbar \Omega }}{2} + \frac{{\hbar \omega }}{2}\left( {1 + 2\frac{\omega }{\Omega }} \right)
\end{equation}
which brings together both non-adiabatic corrections to the Berry phase. While both are important for Fig. 1c,d of the main text, only the spinor part plays a role for the wavepacket trajectory (Fig. 4 of the main text), since the latter is determined by the Berry curvature, as we will show below.

\subsection{Semiclassical equations of motion}

In 1999, Niu and Sundaram derived the semiclassical equations of motion of an electron wavepacket in Bloch bands \cite{Sundaram1999}. Their approach allowed to confirm the anomalous Hall effect predicted by Karplus and Luttinger \cite{Luttinger1954} and to express this effect in terms of Berry curvature. Following this approach, we derive our corrected equations up to the second order in external constant force for a two-band system using perturbation theory. We consider the case of geodesic trajectory of the wavepacket, corresponding to the evolution of the effective field on the equator of the Bloch sphere. We show that in this configuration, all first-order and most of the second-order corrections to the semiclassical equations of motion are zero, except the particular correction due to the difference of the metric on the equator and at the equilibrium deviation angle of the spin. 

We define the two-band wavepacket wavefunction with a well defined center of mass position $\mathbf{r_c}$ by:
\begin{equation}
\ket{W}=\int a(\mathbf{k},t)(f_0 \ket{\psi_{0\mathbf{k}}}+f_1 \ket{\psi_{1\mathbf{k}}}) d\mathbf{k}
\end{equation}
Where $a(\mathbf{k})$ is the distribution of the wavepacket in momentum space centred at $\mathbf{k}=\mathbf{k_c}$  and $\int |a(\mathbf{k})|^2d\mathbf{k}=1$. 
Then, the eigenfunctions of the two bands $\psi_{0,1}$ (which are the aligned and anti-aligned states in the effective field formalism) can be decomposed on the basis of the Bloch waves:
\begin{equation}
\ket{\psi_{i\mathbf{k}}}=e^{i \mathbf{k.r}}\ket{u_{i}} ,~~ |f_0|^2+|f_1|^2=1
\end{equation}
where $\ket{u_i}$ are the Bloch eigenstates.
The effective Lagrangian of the wave packet can then be written as \cite{RevModPhys.82.1959, Sundaram1999,Culcer2005}: 
$\mathcal{L}=\bra{W}(i\frac{d}{d t}-H)\ket{W}$. The Hamiltonian can be expanded around $\mathbf{r_c}$ using gradient correction in external electric potential $V$ (the second order gradient correction is zero since we consider a constant external force):
\begin{eqnarray}
H&=H_c-e(\mathbf{r}-\mathbf{r_c}).\nabla{V(\mathbf{r})}\\
&=H_c-e(\mathbf{r}-\mathbf{r_c}).\mathbf{E}
\end{eqnarray}
$H_c$ is the exact quantum Hamiltonian evaluated at $\mathbf{r_c}$
\begin{eqnarray}
H_c=H_0-eV(\mathbf{r_c})
\end{eqnarray}
where $H_0$ is the unperturbed Hamiltonian ($H_0\ket{\psi_n}=\epsilon_n\ket{\psi_n}$).

Let us look on the first term of the Lagrangian:
\begin{widetext}
\begin{equation}
\bra{W}i\hbar \frac{d}{d t}\ket{W}= i\hbar( f_0^*\frac{d}{d t}f_0+f_1^*\frac{d}{d t}f_1 )+|f_0|^2\mathbf{A}_{00}^t+|f_1|^2\mathbf{A}_{11}^t+f_1^*n_2\mathbf{A}_{01}^t+f_1^*f_0(\mathbf{k_c})\mathbf{A}_{10}^t +\hbar \mathbf{q_c.} \stackrel{.}{\mathbf{r}}_c 
\end{equation}
\end{widetext}
where $n_{i,j}=n_{i,j}(\mathbf{k_c})$. For a general two-level system, we can map the spinor Bloch wave function on the Bloch sphere. Then, we analyze the configuration where the wavepacket is moving on a geodesic in momentum space which means that the effective magnetic field is moving on a great circle (the equator) in the Bloch sphere picture. 
A general wave function can be written as a superposition of the instantaneous eigenstates which are aligned and anti-aligned with an effective field ($\ket{u_{0,1}}=1/\sqrt{2}(e^{-i\phi},\pm 1)^T$).

\begin{equation}
\ket{u}=f_0 \ket{u_0}+f_1 \ket{u_1}=(\cos\theta/2 e^{-i(\phi+\eta)},\sin\theta/2)^T
\end{equation}

We are interested in the mean correction which is computed by considering the equilibrium position ($\theta=\theta_{eq}$, $\eta=0$). In this equilibrium regime, the deviation of the spin from the magnetic field is constant which means that 
\begin{equation}
\frac{df_i}{dt}=0.
\end{equation}
We can hence simplify the previous equation.

Using the expression for the terms involving the Berry connection:
\begin{equation}
\mathbf{A}_{ij}^t=i\bra{u_i}\frac{d}{dt}\ket{u_j}
\end{equation}
one has 
\begin{equation}
\mathbf{A}_{00}^t=\mathbf{A}_{11}^t=\frac{1}{2}\frac{d\phi}{dt}
\end{equation}
and finally we obtain
\begin{widetext}
\begin{equation}
\bra{W}i\hbar \frac{d}{d t}\ket{W}= i\hbar(\bra{u_0}\frac{d}{dt}\ket{u_0}+f_0^*f_1\bra{u_0}\frac{d}{dt}\ket{u_1}+f_1^*f_0\bra{u_1}\frac{d}{dt}\ket{u_0})+\hbar \mathbf{q_c.} \stackrel{.}{\mathbf{r}}_c
\end{equation}
\end{widetext}

The time derivatives in this expression are to be replaced by the derivatives in the parameter space, and an important step to take into account the non-adiabaticity correctly, is to keep in the equation the equilibrium deviation of the spin from the magnetic field (i.e. keep track of the fact that the spin is not on the equator). In order to do this, using the Bloch sphere picture as discussed in the main text, we can rewrite:
\begin{equation}
\bra{u_i}\frac{d}{dt}\ket{u_j}=\bra{u_i}\frac{\partial}{\partial s_{s}}\ket{u_j}\frac{ds_{s}}{dt}=\frac{1}{\sin\theta_{eq}}\bra{u_i}\frac{\partial}{\partial \phi}\ket{u_j}\frac{ds_{s}}{dt}
\end{equation}
where $ds_{s}$ corresponds to the metric of the sphere.
For small non-adiabaticity, the deviation of the spin from the equator is small and we can expand the correction in Taylor series:
\begin{equation}
 1/\sin\theta_{eq}=1/\sin(\pi/2-\zeta_{eq})\approx 1+\zeta_{eq}^2/2
\end{equation}

Moreover, we can express $\omega$ in term of the \textit{quantum distance} $ds_q$:
$\omega=\frac{d\phi}{dt}=2\frac{ds_q}{dt}=2\sqrt{g_{ij}\frac{dk_i}{dt}\frac{dk_j}{dt}}$ .
Using these developments, and the fact that here $\hbar \Omega =\epsilon_0-\epsilon_1$ we can write the effective Lagrangian of the wavepacket in momentum space as:

\begin{widetext}
\[\mathcal{L} = \hbar \mathbf{k_c.} \stackrel{.}{\mathbf{r}}_c +\hbar(1+\frac{ 2 \hbar^2 g_{ij}}{(\epsilon_0-\epsilon_1)^2} \frac{\partial k_i}{\partial t}\frac{\partial k_j}{\partial t}) \left\{ {{}{{\bf{A}}_{00}} + {f_0}^*{f_1}{{\bf{A}}_{01}} + {f_1}^*{f_0}{{\bf{A}}_{10}}} \right\}.\frac{{\partial {{\bf{k}}_{\bf{c}}}}}{{\partial t}} - {\left| {{f_0}} \right|^2}{\epsilon_0}\left( {{{\bf{k}}_{\bf{c}}}} \right) - {\left| {{f_1}} \right|^2}{\epsilon_1}\left( {{{\bf{k}}_{\bf{c}}}} \right) + eV \left( {{{\bf{r}}_{\bf{c}}}} \right)\]
\end{widetext}
We can see that the correction term which takes into account the non-adiabaticity appears in the second order of the momentum derivative. As a consequence, we need to use perturbation theory up to the second order in the external force in order to make the computation self-consistent.
These results are well known: 

\begin{widetext}
%${n_0} = 1 - \frac{{\left\langle {{u_0}} \right|{\bf{E}}.{{\bf{r}}_c}\left| {{u_1}} \right\rangle \left\langle {{u_1}} \right|{\bf{E}}.{{\bf{r}}_c}\left| {{u_0}} \right\rangle }}{{2{{\left( {{\epsilon_0} - {\epsilon_1}} \right)}^2}}}$ ~~~~ ${n_1} = \frac{{\left\langle {{u_0}} \right|{\bf{E}}.{{\bf{r}}_c}\left| {{u_1}} \right\rangle }}{{\left( {{\epsilon_0} - {\epsilon_1}} \right)}} - \frac{{\left\langle {{u_1}} \right|{\bf{E}}.{{\bf{r}}_c}\left| {{u_0}} \right\rangle \left\langle {{u_0}} \right|{\bf{E}}.{{\bf{r}}_c}\left| {{u_0}} \right\rangle }}{{{{\left( {{\epsilon_0} - {\epsilon_1}} \right)}^2}}} + \frac{{\left\langle {{u_1}} \right|{\bf{E}}.{{\bf{r}}_c}\left| {{u_1}} \right\rangle \left\langle {{u_1}} \right|{\bf{E}}.{{\bf{r}}_c}\left| {{u_0}} \right\rangle }}{{{{\left( {{\epsilon_0} - {\epsilon_1}} \right)}^2}}}$
\begin{eqnarray}
f_0=1-\frac{e^2}{2}\frac{\bra{u_0}\mathbf{E.r_c}\ket{u_1}\bra{u_1}\mathbf{E.r_c}\ket{u_0}}{(\epsilon_0-\epsilon_1)^2}\\
f_1=-e\frac{\bra{u_0}\mathbf{E.r_c}\ket{u_1}}{(\epsilon_0-\epsilon_1)}-e^2\frac{\bra{u_1}\mathbf{E.r_c}\ket{u_0}\bra{u_0}\mathbf{E.r_c}\ket{u_0}}{(\epsilon_0-\epsilon_1)^2}+e^2\frac{\bra{u_1}\mathbf{E.r_c}\ket{u_1}\bra{u_1}\mathbf{E.r_c}\ket{u_0}}{(\epsilon_0-\epsilon_1)^2}
\end{eqnarray} 
\end{widetext} 
 where the scalar product can be expressed in terms of the intra-band and inter-band Berry connexions $\mathbf{A}_{ij}$:
  
\begin{equation}
\bra{u_i}\mathbf{E.r_c}\ket{u_j}=\mathbf{E}.\bra{u_i}i\frac{\partial}{\partial \mathbf{k_c}}\ket{u_j}=\mathbf{E}.\mathbf{A}_{ij}
\end{equation}
where $\mathbf{r}_{\mathbf{c}}$ and $\mathbf{q}_{\mathbf{c}}$ are the position and momentum of the wavepacket center of mass. 
$\epsilon_i$ are the energies of the two bands. $E$ is the constant electric field ($\mathbf{E}=-\frac{\partial V}{\partial \mathbf{r_c}}$).
Considering the electric field in the $x$ direction $\mathbf{E}=E \mathbf{e_x}$, with no loss of generality, and using the identity \cite{Gao2015,Piechon2016}: 

\begin{equation}
g_{\alpha \beta}=\frac{1}{2}(\bra{u_0}i\frac{\partial}{\partial k_\alpha}\ket{u_1}\bra{u_1}i\frac{\partial}{\partial k_\beta}\ket{u_0}+h.c.)
\end{equation}
we can simplify the terms proportional to the Berry connection:
\begin{widetext}
\begin{equation}
\left\{ {{}{{\bf{A}}_{00}} + {f_0}^*{f_1}{{\bf{A}}_{01}} + {f_1}^*{f_0}{{\bf{A}}_{10}}} \right\}\mathbf{.} \frac{\partial\mathbf{k}}{\partial t}=A_{00}^x \stackrel{.}{k}_x+A_{00}^y \stackrel{.}{k}_y- 2 e \frac{E_x(g_{xx}\stackrel{.}{k}_x+g_{xy}\stackrel{.}{k}_y)}{(\epsilon_0-\epsilon_1)}+2e^2 \frac{E_x^2(A_{11}^x-A_{00}^x)}{(\epsilon_0-\epsilon_1)^2}(g_{xx}\stackrel{.}{k}_x+g_{xy}\stackrel{.}{k}_y)
\end{equation}
\end{widetext}
The second order correction in electric field due to time-independent perturbation theory vanishes in this term because $A_{00}^x=\bra{u_0}i\frac{\partial}{\partial k_x}\ket{u_0}=A_{11}^x$ on the geodesics. The first-order correction is the correction found by Gao and Niu in \cite{Gao2014,Gao2015} recently. This term does not appear in the equations of motion when the wavepacket is following a geodesic in momentum space (see the section on geodesic trajectories).

We can define the effective Hamiltonian $\mathcal{H}=\sum_i {\stackrel{.}{q}_i \frac{\partial \mathcal{L}}{\partial \stackrel{.}{q}_i}} - \mathcal{L} $ and then deduce the equations of motion:
%\begin{widetext}
%\[L = \frac{{d\gamma }}{{dt}} - {{\bf{r}}_{\bf{c}}}.\frac{{\partial {{\bf{q}}_{\bf{c}}}}}{{\partial t}} + \left\{ {{}{{\bf{A}}_{11}} + {n_1}^*{n_2}{{\bf{A}}_{12}} + {n_2}^*{n_1}{{\bf{A}}_{21}}} \right\}.\frac{{\partial {{\bf{q}}_{\bf{c}}}}}{{\partial t}} - {\left| {{n_1}} \right|^2}{E_1}\left( {{{\bf{q}}_{\bf{c}}}} \right) - {\left| {{n_2}} \right|^2}{E_2}\left( {{{\bf{q}}_{\bf{c}}}} \right) + e\phi \left( {{{\bf{r}}_{\bf{c}}}} \right)\]
%\end{widetext}

\begin{widetext}
\[\mathcal{H} ={\left| {{f_0}} \right|^2}{\epsilon_0}\left( {{{\bf{k}}_{\bf{c}}}} \right) + {\left| {{f_1}} \right|^2}{\epsilon_1}\left( {{{\bf{k}}_{\bf{c}}}} \right) + eEx-\hbar(1+\frac{2\hbar^2g_{xx}}{(\epsilon_0-\epsilon_1)^2}\stackrel{.}{k}_x^2 ) \left\{A_{00}^x \stackrel{.}{k}_x+A_{00}^y \stackrel{.}{k}_y \right\} \]
\end{widetext}
Using the Hamilton equations, with $p_i=\hbar k_i$ 
\begin{equation}\frac{dp_i}{dt}=-\frac{\partial \mathcal{H}}{\partial q_i} ,~~ \frac{dq_i}{dt}=\frac{\partial \mathcal{H}}{\partial p_i}
\end{equation} 
we can derive the gauge invariant equations of motion:

\begin{eqnarray}
\hbar\frac{\partial\textbf{k}}{\partial t}=\textbf{F}\\
\hbar \frac{\partial\textbf{r}}{\partial t}= \frac{\partial \tilde{\epsilon}}{\partial \textbf{k}}  - \hbar \frac{\partial\textbf{k}}{\partial t} \times \mathbf{B}\left(1+\frac{2\hbar^2 g_{kk}}{(\epsilon_0-\epsilon_1)^2}\left(\frac{\partial\textbf{k}}{\partial t}\right)^2\right)
\label{semiclassical}
\end{eqnarray}
with the Berry curvature and the corrected energy:
\begin{equation}
\mathbf{B}=\nabla_\mathbf{k}\times\mathbf{A_{00}}
\end{equation}
\begin{equation}
\tilde{\epsilon}=|f_0|^2\epsilon_0+|f_1|^2\epsilon_1
\end{equation}

Here we use the notation $\mathbf{F}$ for a general external constant force. Indeed, these equations are valid for a wavepacket of neutral particles in a gradient potential $\mathbf{F}=-\nabla U$, which is the applied case studied in the main text. 
Equation (34) corresponds to equation (9) of the main text which is written in term of the QGT components and with $\epsilon_0-\epsilon_1=\hbar\Omega$.
This is the central result presented and verified by numerical simulations in the main text.

\subsection{First-order corrections on geodesics} 
First-order non-adiabatic corrections to the semiclassical equations have been found in the previous works \cite{Gao2014,Gao2015,Gao2016}. Here, we show that these corrections from the semiclassical equations of motion disappear if the system evolves along the geodesic trajectories (for example, accelerated from zero average initial velocity by a constant force, as in the Hall effect configuration).

We want to analyze the first-order contribution of the real part of the quantum geometric tensor to the trajectory along the geodesic lines in the parameter space. This contribution to the dynamical equation reads:
\begin{equation}
{\bf{\dot r}} =  {\bf{\dot k}} \times{\bf{B }}' 
\end{equation}
where the first order correction of the Berry curvature is
\begin{equation}
{\bf{B}}' = \nabla  \times \mathord{\buildrel{\lower3pt\hbox{$\scriptscriptstyle\leftrightarrow$}} 
\over G} {\bf{E}}
\end{equation}
Here, $\mathord{\buildrel{\lower3pt\hbox{$\scriptscriptstyle\leftrightarrow$}} 
\over G}$  is a tensor proportional to the quantum metric tensor of a single band in the case of two bands.
Let us consider a particle accelerated from $k=0$ in the $X$ direction by a field along the $X$ axis  ${\bf{E}} = {E_x}{{\bf{e}}_x}$. The contribution to the trajectory in the transverse direction is given by
\begin{equation}
\dot y = B {'_z}{\dot k_x}
\end{equation}
where
\begin{equation}
B{'_z} = \frac{\partial }{{\partial {k_x}}}{G_{yx}}{E_x} - \frac{\partial }{{\partial {k_y}}}{G_{xx}}{E_x}
\label{mtenseq}
\end{equation}
Other terms (like $G_{yy}E_y$) disappear because  $E_y=0$. Note, that the indices of the tensor correspond to the coordinates in the reciprocal space, but are written as $x,y$  to save space. The "real space" component of the electric field actually represents the time derivative of the wave vector as well: ${E_x} = d{k_x}/dt$.
We need to analyze each of these two terms. First, we show that the off-diagonal components of the metric tensor are zero when the basis vectors are chosen tangential and perpendicular to the geodesic curve, along which the system propagates. The first term is therefore zero. Then we show that the second term is zero as well. The first order contribution thus disappears.
\subsubsection{Metric tensor along geodesic lines}
The length of a parametric curve in a space with a metric tensor $g_{ij}$  is given by:
\begin{equation}
L = \int {\sqrt {\sum\limits_{i,j} {{g_{ij}}\frac{{d{x_i}}}{{dt}}\frac{{d{x_j}}}{{dt}}} } dt}
\end{equation}
The length of a small part of this curve is simply
\begin{equation}
dL = \sqrt {\sum\limits_{i,j} {{g_{ij}}\frac{{d{x_i}}}{{dt}}\frac{{d{x_j}}}{{dt}}} } dt
\end{equation}
We choose the basis vector $x_i$  to be tangential to the trajectory and $x_j$  to be normal to the trajectory. The latter means that $dx_j/dt=0$. Indeed, if we imagine that a certain object is propagating along the $X$ axis, it necessarily has zero velocity in the perpendicular direction, otherwise it would not be propagating along $X$. In this basis, we have therefore a simplified expression for the length of the curve:
\begin{equation}
dL = \sqrt {{g_{ii}}} d{x_i}
\end{equation}
which is quite natural. We suppose that this parametric curve is a geodesic one, which means that it has the smallest possible length.
Now, let us consider a curve which deviates slightly from the geodesic curve: $dx_j/dt$  is not zero, but  much smaller than  $dx_i/dt$. The fact that the geodesic curve has the smallest possible length means that  $dL$ is minimized with respect to $dx_j$ , and therefore its derivative should be zero:
\begin{equation}
\frac{{\partial L}}{{\partial {x_j}}} = 0
\end{equation}
The length of this curve now reads
\begin{equation}
dL = \sqrt {{g_{ii}}{{\left( {\frac{{d{x_i}}}{{dt}}} \right)}^2} + {g_{jj}}{{\left( {\frac{{d{x_j}}}{{dt}}} \right)}^2} + 2{g_{ij}}\frac{{d{x_i}}}{{dt}}\frac{{d{x_j}}}{{dt}}} dt
\end{equation}
Neglecting the second order term and expanding the square root in series of  $dx_j/dt$, we obtain
\begin{equation}
dL \approx \sqrt {{g_{ii}}} d{x_i} + \frac{{{g_{ij}}}}{{\sqrt {{g_{ii}}} }}d{x_j}
\end{equation}
We now see that the geodesic nature of the curve
\begin{equation}
\frac{{\partial L}}{{\partial {x_j}}} = \frac{{{g_{ij}}}}{{\sqrt {{g_{ii}}} }} = 0
\end{equation}
is equivalent to the diagonality of the metric tensor in tangential/normal basis:
\begin{equation}
{g_{ij}} = 0
\end{equation}
and therefore the first contribution of Eq.  \eqref{mtenseq} is zero.

\subsubsection{Transverse derivative of the metric tensor}
We need to analyze the second term given by
\begin{equation}
\frac{\partial }{{\partial {k_y}}}{g_{xx}}{E_x}
\end{equation}
While in general, the diagonal term $g_{xx}$  can, of course, depend on $k_y$, the first order derivative is zero on the geodesic line, as we show below. Indeed, on one hand we have
\begin{equation}
dL = \sqrt {{g_{xx}}} d{k_x}
\end{equation}
And on the other hand, the minimal length of the geodesic requires that
\begin{equation}
\frac{{\partial L}}{{\partial {k_y}}} = 0
\end{equation}
Thus,
\begin{equation}
\frac{1}{{2\sqrt {{g_{xx}}} }}\frac{{\partial {g_{xx}}}}{{\partial {k_y}}} = 0
\end{equation}
And finally
\begin{equation}
\frac{{\partial {g_{xx}}}}{{\partial {k_y}}} = 0
\end{equation}
This contribution to Eq. \eqref{mtenseq} is also zero along a geodesic line.

\subsection{Polariton non-adiabatic evolution}
Now, we can apply the new equations to determine the non-adiabaticity of a polariton wavepacket, accelerated in combined TE-TM and Zeeman fields. In this case the total magnetic field energy reads:

\begin{equation}
\hbar\Omega  = \sqrt {{\Delta ^2} + {{\left( {\beta {k^2}} \right)}^2}} 
\end{equation}

When the polaritons propagate in a constant gradient, their wave vector increases linearly with time: $k=F t/\hbar$, where the force can be obtained for example by applying a gradient $F=-\nabla U$.
The angular velocity of the rotation of the total magnetic field can be obtained as:

\begin{equation}
\omega  = \frac{{d\varphi }}{{dt}} = \frac{{2ds}}{{dt}} = 2\frac{{ds}}{{dk}}\frac{{dk}}{{dt}} = \frac{{2F}}{\hbar }\sqrt {{g_{kk}}} 
\end{equation}
which gives
\[\omega  = \frac{{2F}}{\hbar }\frac{{\Delta \beta k}}{{\left( {{\Delta ^2} + {{\left( {\beta {k^2}} \right)}^2}} \right)}}\]

The wavepacket is created at $k=0$ and accelerated towards $k=\infty$. Its non-adiabaticity can be obtained as 
\begin{equation}
{f_{NA}}\left( k \right) =\frac{1}{4} \min \left( {\theta _{eq}^2,\theta _0^2} \right)
\end{equation}
where
\begin{equation}
\theta _{eq}^2 = {\left( {\frac{{2F}}{\hbar }} \right)^2}\frac{{{\Delta ^2}{\beta ^2}{k^2}}}{{{{\left( {{\Delta ^2} + {{\left( {\beta {k^2}} \right)}^2}} \right)}^3}}}
\end{equation}
and 
\begin{equation}
\theta _0^2 = \frac{{{\beta ^2}{F^2}}}{{{\Delta ^6}{\hbar ^2}}}
\label{nares}
\end{equation}
The latter gives us the residual non-adiabaticity measured at large wavevectors and times. We see that it is not exponential in $F$, but linear, because the configuration of the experiment is different from that of Landau-Zener: instead of slowly increasing the perturbation with time (when we approach the "coupling region" from $-\infty$), we suddenly  turn it on at $t=0$.

Since the residual non-adiabaticity corresponds to precession around $\theta_{eq}=0$, it does not contribute to the extra phase or to the wavepacket trajectory, and does not appear in the dynamical equations in the main text. For realistic parameters of microcavities, the residual non-adiabaticity looks to be inacessible, requiring long propagation distances. Therefore, the expression for the non-adiabaticity is simplified, and one can write it as:

\begin{equation}
{f_{NA}}\left( k \right) = \frac{{{F^2}{g_{kk}}}}{{{\hbar^2\Omega ^2}}}
\label{pnaqgt}
\end{equation}

We stress that the non-adiabatic fraction and the corresponding correction are calculated from the adiabatic metric.

\subsection{Experimental measurement of non-adiabaticity}
In the precise configuration we consider, the non-adiabaticity of the wavepacket can be measured via the diagonal polarization degree. For this, one should orient the cavity in such a way that the wedge points in the $X$ direction. Then, the "horizontal" polarization should be chosen parallel to this $X$ axis. In this case, the adiabatic evolution means that the spin rotates from the $Z$ axis (circular eigenstate at $k=0$) to the $X$ pseudospin axis (horizonal polarization), while the deviation from this rotation is necessarily towards the $Y$ pseudospin axis (diagonal polarization).

\begin{figure}[tbp]
\includegraphics[scale=0.7]{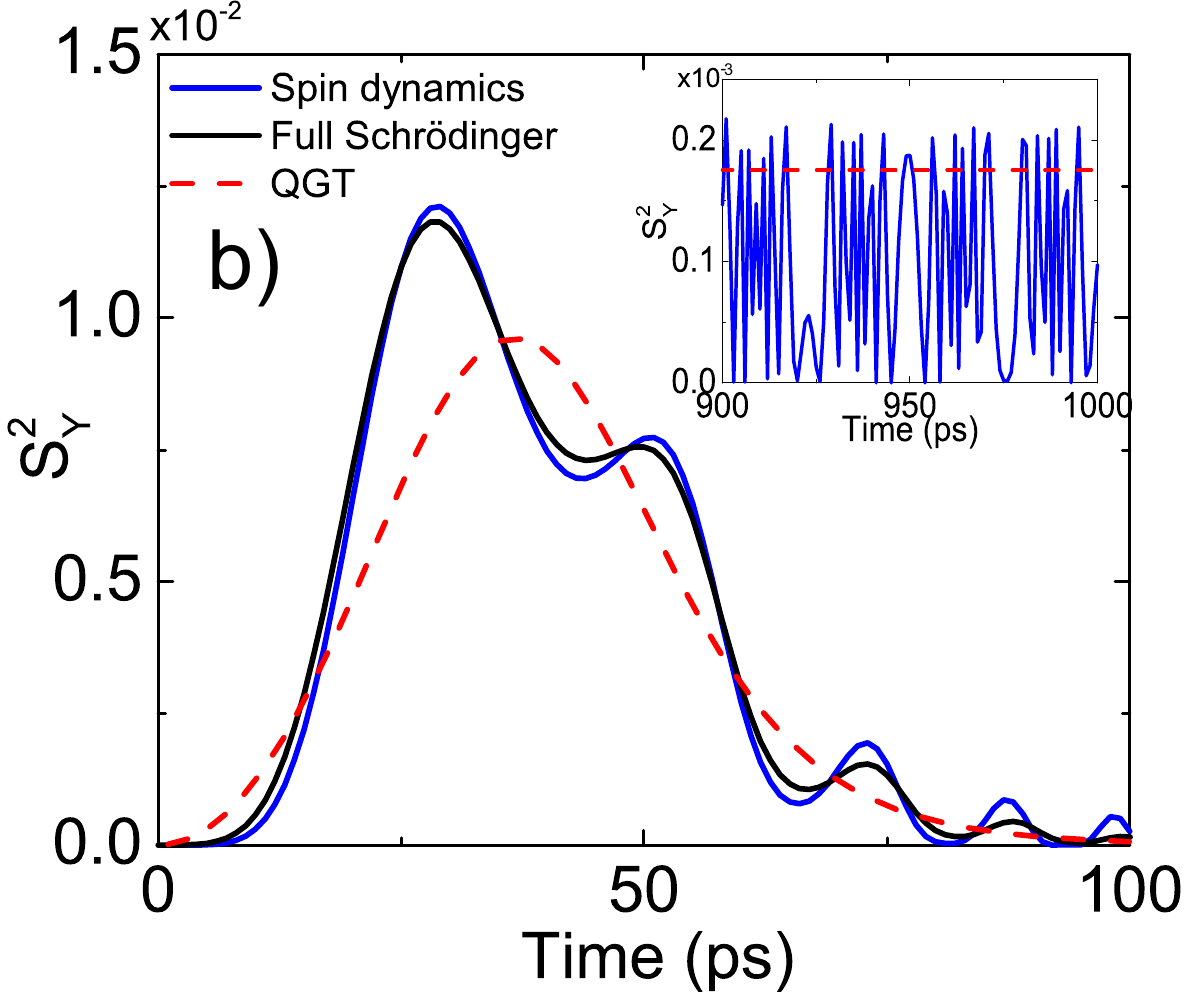}
\caption{\label{figdisp} (Color online) The nonadiabaticity ($S_Y^2$), calculated using the full Schrodinger equation (black), the spin dynamics equation (blue), and the QGT (red dashed).}
\end{figure}

The three pseudospin components can be extracted from the spatially-integrated intensities measured in 3 pairs of orthogonal polarizations:
\begin{eqnarray}
S_x = \frac{{{I_H} - {I_V}}}{{{I_H} + {I_V}}}, ~
S_y = \frac{{{I_D} - {I_A}}}{{{I_D} + {I_A}}}, ~
S_z = \frac{{{I_ + } - {I_ - }}}{{{I_ + } + {I_ - }}}
\end{eqnarray}

\begin{figure}[tbp]
\includegraphics[scale=0.6]{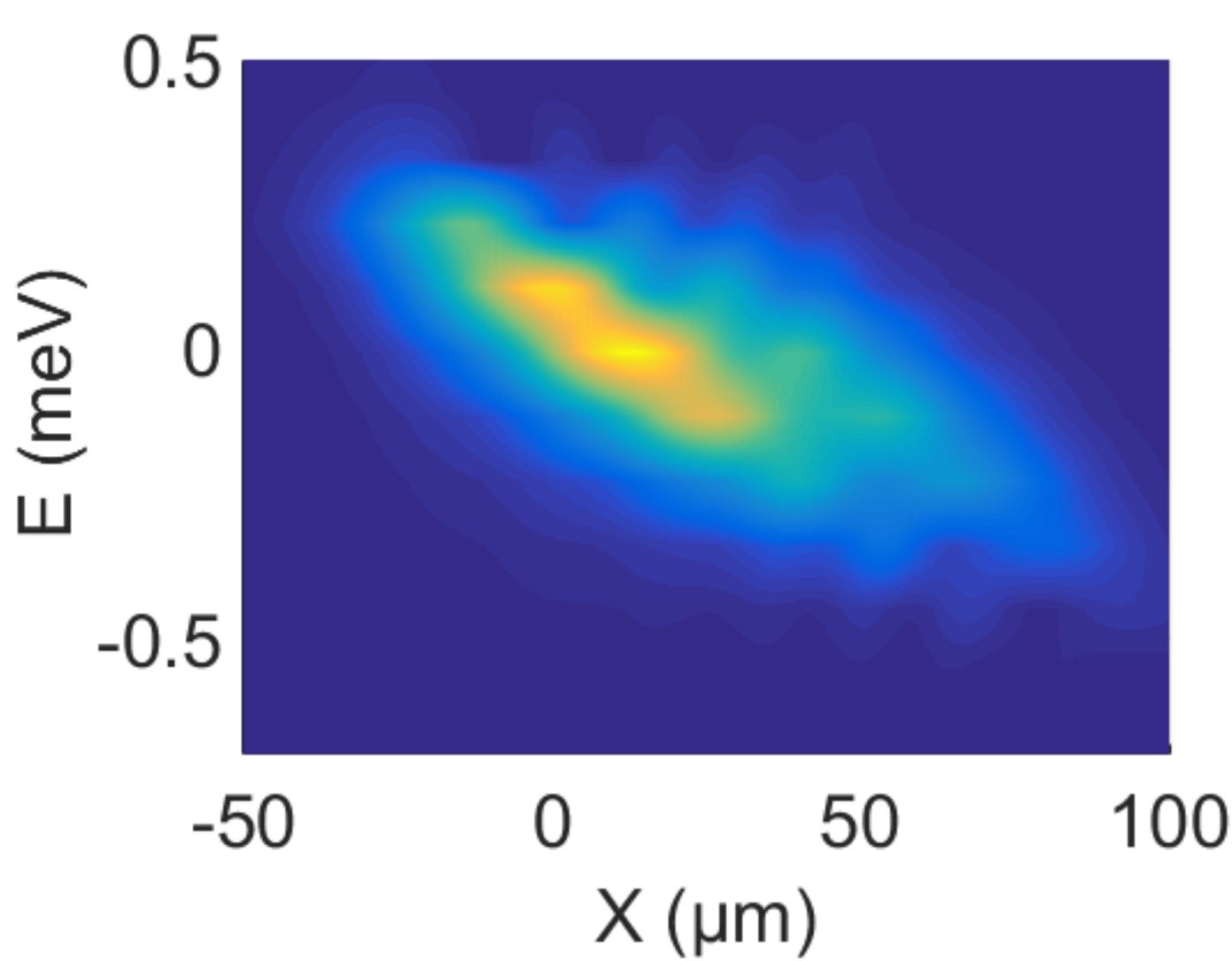}
\caption{\label{figEX} (Color online) Energy-resolved spatial image of emitted intensity $|\psi(x,E)|^2$ for a time window.}
\end{figure}

Figure 1 shows the agreement between the nonadiabaticity, given by the $S_Y^2$ pseudospin component, calculated by solving full Schrodinger equations (black), the simple spin dynamics equation, Eq. \eqref{dyneq} (blue), and given by the component of the QGT, Eq. \eqref{pnaqgt} (red dashed). We see that the latter captures very well the behavior of the full system, and that the spin dynamics equation for small angles is sufficiently precise.  The figure is plotted for $\beta=0.14$ meV/$\mu$m$^{-2}$. The inset shows the residual non-adiabatic fraction $f_{NA,0}$ (blue) and the analytical value given by Eq. \eqref{nares} (red dashed).

Another option to measure the non-adiabaticity could be to analyze the spectrum of the wavefunction, because the energy of the ground and excited states is different. However, this approach encounters difficulties in the experimental setting that we propose because of the spatial gradient which accelerates particles. Indeed, different points of the wavepacket have different potential energy at a given moment of time. Moreover, since the wavepacket propagates down the slope, its energy changes quite rapidly over time. Still, it is possible to use a time window and see the excited state in the energetically and spatially resolved emission intensity $|\psi(x,E)|^2$, as shown in Fig. S\ref{figEX}.

\subsection{Wavepacket trajectory}
The anomalous Hall effect with Rashba-type SOC (linear in $k$) is quite well-known, and the corresponding Berry curvature and trajectory (without non-adiabatic corrections) have been calculated before. This configuration corresponds both to Dirac Hamiltonian describing an electron in vacuum \cite{Chang2008} and to electrons in 2D materials (like Transitional Metal Dichalcogenides), which obey to an effective Dirac Hamiltonian \cite{Xu2014}.

At the same time, the case of $k^2$ TE-TM SOC with double winding has not been studied, although it is quite important for photonics. In this section, we present the detailed results concerning the trajectory of an accelerated wavepacket (under the effect of a force $F$) in presence of the TE-TM SOC and a constant magnetic field providing Zeeman splitting, solving equations (6) and (9) of the main text.

The analytical solution of eq. (6) which does not include the non-adiabatic correction reads:
\begin{equation}
x\left( t \right) = \frac{{F{t^2}}}{{2m}} - \frac{{\sqrt {{\Delta ^2} + \frac{{{\beta ^2}{F^4}{t^4}}}{{{\hbar ^4}}}} }}{F} + \frac{\Delta }{F}
\end{equation}
and
\begin{widetext}
\begin{equation}
y\left( t \right) = \frac{{{F^3}\hbar {t^3}{\beta ^2}\sqrt {{\Delta ^2} + {F^4}{t^4}{\beta ^2}/{\hbar ^4}} \left( {3 - \left( {1 + \frac{{{F^4}{t^4}{\beta ^2}}}{{{\Delta ^2}{\hbar ^4}}}} \right){\,_2}{F_1}\left( {1,\frac{5}{4},\frac{7}{4}, - \frac{{{F^4}{t^4}{\beta ^2}}}{{{\Delta ^2}{\hbar ^4}}}} \right)} \right)}}{{3\left( {{\Delta ^3}{\hbar ^4} + \Delta {F^4}{t^4}{\beta ^2}} \right)}}
\end{equation}
\end{widetext}

The maximal shift of the wavepacket in the $Y$ direction at $t\to\infty$ is given by:
\begin{equation}
y_{\infty}=\frac{\sqrt{\beta}}{\sqrt{\Delta}}\frac{\Gamma^2\left(\frac{3}{4}\right)}{\sqrt{\pi}}
\end{equation}
We see that increasing the lateral shift requires increasing the TE-TM splitting $\beta$ or decreasing the magnetic field $\Delta$.

The importance non-adiabatic correction can be estimated by the maximal value of the non-adiabatic fraction $f_{NA}$ behaves as:
\begin{equation}
f_{NA,max}\propto\frac{\beta}{\Delta^3}
\end{equation}
To reduce the maximal non-adiabaticity, it is therefore important to have a sufficiently large Zeeman splitting $\Delta$. To have a maximal displacement and a good adiabaticity at the same time, it is better to increase the TE-TM SOC $\beta$ instead of reducing the magnetic field $\Delta$, because the non-adiabaticity behaves as $1/\Delta^3$.

The analytical solution for the equation (9) of the main text including the non-adiabatic correction can also be found:
\begin{widetext}
\begin{equation}
y\left( t \right) = \frac{{{F^3}{\beta ^3}\left( {\Delta t\left( {15 + 2{\Delta ^2}\left( {70{t^2} + \frac{{ - 15{\Delta ^4}{\hbar ^{12}} + 8{\Delta ^2}{F^4}{\hbar ^8}{t^4}{\beta ^2} + 3{F^8}{\hbar ^4}{t^8}{\beta ^4}}}{{{{\left( {{\Delta ^2}{\hbar ^4} + {F^4}{t^4}{\beta ^2}} \right)}^3}}}} \right)} \right) + \frac{{5{{\left( { - 1} \right)}^{3/4}}\Delta \hbar \sqrt {\frac{\Delta }{\beta } + \frac{{{F^4}{t^4}\beta }}{{\Delta {\hbar ^4}}}} \nu}}{{{F^3}\beta }}} \right)}}{{140{\Delta ^4}{\hbar ^3}\sqrt {{\Delta ^2} + {F^4}{t^4}{\beta ^2}/{\hbar ^4}} }}
\end{equation}
\end{widetext}
where
\begin{widetext}
\begin{equation}
\nu  = 28{\Delta ^3}{\hbar ^2}{\mathop{\rm E}\nolimits} \left( {i{\mathop{\rm arcsinh}\nolimits} \left( {\frac{{Ft\sqrt {\frac{{i\beta }}{\Delta }} }}{\hbar }} \right), - 1} \right) + \left( { - 28{\Delta ^3}{\hbar ^2} + 3i{F^2}\beta } \right){\mathop{\rm F}\nolimits} \left( {i{\mathop{\rm arcsinh}\nolimits} \left( {\frac{{Ft\sqrt {\frac{{i\beta }}{\Delta }} }}{\hbar }} \right), - 1} \right)
\end{equation}
\end{widetext}
where $\mathrm{E}$ and $\mathrm{F}$ are the elliptic integrals of the second kind. Unfortunately, we did not manage to find the limit of this expression for $t\to\infty$. 

From the practical point of view, the above analytical expressions for $y(t)$ appear quite cumbersome, and we advise the direct numerical solution of equations (6) and (9), because the calculation time of the hypergeometric function and the elliptic integrals is quite comparable or even longer than the direct solution of the equations, while the results are identical.

\subsection{Supplementary Video}

The supplementary movie \textbf{movie.avi} (\url{https://youtu.be/xYaD6ql9opg}) shows the motion of a polariton wavepacket accelerated by a potential gradient. The center of mass of the wavepacket is marked with a white cross, which allows to see its lateral deviation, occurring because of the Berry curvature, as expected from the dynamical equations.

\end{document}